\documentclass[revsymb,twocolumn,prb]{revtex4}%
\usepackage{graphicx}
\usepackage{bm}
\usepackage{slashed}
\usepackage{amsmath}
\usepackage{amsfonts}
\usepackage{amssymb}
\usepackage{lipsum}%
\providecommand{\U}[1]{\protect\rule{.1in}{.1in}}
\providecommand{\U}[1]{\protect\rule{.1in}{.1in}}
\newcommand{\be}{\begin{equation}}
\newcommand{\en}{\end{equation}}
\begin{document}
\title{Functional renormalization-group approach to the Pokrovsky-Talapov model via
modified massive Thirring fermion model}
\author{P.A. Nosov,$^{1}$ Jun-ichiro Kishine,$^{2,3}$, A.S. Ovchinnikov,$^{1,4}$ and
I. Proskurin$^{1,2,3}$}
\affiliation{$^{1}$ Institute of Natural Science, Ural Federal University, Ekaterinburg
620002, Russia }
\affiliation{$^{2}$ Center for Chiral Science, Hiroshima University, Higashi-Hiroshima,
Hiroshima 739-8526, Japan}
\affiliation{$^{3}$ Division of Natural and Environmental Sciences, The Open University of
Japan, Chiba 261-8586, Japan}
\affiliation{$^{4}$ Institute of Metal Physics, Ural Division, Russian Academy of Sciences,
Ekaterinburg 620219, Russia }
\date{\today }

\begin{abstract}
A possibility of the topological Kosterlitz-Thouless~(KT) transition in the
Pokrovsky-Talapov~(PT) model is investigated by using the functional
renormalization-group (RG) approach by Wetterich. Our main finding is that the
nonzero misfit parameter of the model, which can be related with the linear
gradient term (Dzyaloshinsky-Moriya interaction), makes such a transition
impossible, what contradicts the previous consideration of this problem by
non-perturbative RG methods. To support the conclusion the initial PT model is
reformulated in terms of the 2D theory of relativistic fermions using an
analogy between the 2D sine-Gordon and the massive Thirring models. In the new
formalism the misfit parameter corresponds to an effective gauge field that
enables to include it in the RG procedure on an equal footing with the other
parameters of the theory. The Wetterich equation is applied to obtain flow
equations for the parameters of the new fermionic action. We demonstrate that
these equations reproduce the KT type of behavior if the misfit parameter is
zero. However, any small nonzero value of the quantity rules out a possibility
of the KT transition. To confirm the finding we develop a description of the
problem in terms of the 2D Coulomb gas model. Within the approach the
breakdown of the KT scenario gains a transparent meaning, the misfit gives
rise to an effective in-plane electric field that prevents a formation of
bound vortex-antivortex pairs.

\end{abstract}

\pacs{Valid PACS appear here}
\maketitle

\section{Introduction}

The sine-Gordon (SG) model with a misfit parameter has been discussed in the
context of the commensurate-incommensurate transition (CIT) of adsorbates on a
periodic potential \cite{Pokrovsky1979,Talapov1980} and is now referred to as
the Pokrovsky-Talapov model. Besides adsorbed atoms on a surface it can be
applied to a variety of commensurate-incommensurate systems, including, for
instance, bilayer quantum-Hall junctions with an in-plane field
\cite{Yang1996}, superconducting films \cite{Besseling2003}, cold atoms
\cite{Blucher2003} and fermionic atoms \cite{Molina2007} in an optical lattice
and graphene \cite{Woods2014}.

One noteworthy feature of the PT model is that it predicts the ground state as
a soliton lattice, which satisfies the time-independent sine-Gordon equation
and describes repeatedly spaced solitons \cite{Bak1982}. For example, for
adsorbed layers on crystal surfaces the substrate provides a periodic
potential relief and the ground state consists of large domains, commensurate
with a substrate lattice constant, separated by regions of bad fit, which are
known as misfit dislocations or domain walls. Likewise, the PT model arises
naturally in the description of ground state properties of the chiral
helimagnets \cite{Dzyaloshinskii1964}, where the misfit parameter can be
attribued to the Dzyaloshinsky-Moriya interaction
\cite{Heurich2003,Kishine2014}. In the latter case, the incommensurate phase
is composed of regions of forced ferromagnetic ordering, caused by an applied
magnetic field, divided by restricted regions of $2\pi$ spin twists. Important
trigger to accelerate systematic studies of the magnetic soliton lattice was
the experiment in which an actual image of the order was observed using
Lorentz microscopy \cite{Togawa2012}. As a result, an essential progress is
being made in understanding of properties of the magnetic soliton lattice in
thin films of chiral helimagnets \cite{Togawa2015,Goncalves2017}.
Notwithstanding the applicability of the PT model to describe properties of
the ground state a question mark still hangs over its ability to explain phase
transitions in the materials. The reason for that are different
order-parameter spaces, $U(1)$, where a direction of magnetization is defined
by a single polar angle, and $SO(3)/U(1)$, where a magnetization is determined
by two polar angles, for the PT-model and the model of the chiral
helimagnet\cite{Kishine2015}, respectively.

For adsorbates on a surface the commensurate-incommensurate phase transition
occurs upon heating, where the commensurate phase undergoes a transition into
an incommensurate relative to the substrate floating solid. According to the
PT theory, key features related to the CIT caused by dynamics of domain walls
which meander entropically and interact with each other. Numerous examples of
such transitions have been reported \cite{Lawley1989}, save for magnetic
systems. Further clarification of details of the CIT demonstrated that the
incommensurate solid becomes unstable to the formation of dislocations leading
to the Kosterlitz-Thouless transition to a fluid state \cite{Coppersmith1981}.
At low temperatures, the KT dislocations are bounded in pairs and the
commensurate phase is maintained. The view is nothing but the KT theory for
dislocation melting \cite{Kosterlitz1973}.

A Wilsonian-type renormalization group analysis of the
commensurate-incommensurate transition in the PT-model, where
incommensurability was considered as a new renormalized parameter, has been
performed in Refs \cite{Puga1982,Horowitz1983}. To describe instability of the
incommensurate phase to presence of free dislocations a free energy equivalent
to an anisotropic XY-model was introduced \cite{Coppersmith1981}. It belongs
to the same universality class as the isotropic version of the model
\cite{Berezinskii1971,Kosterlitz1974,Jose1977}.

A non-perturbative analysis suggested much later by Lazarides et al.
\cite{Lazarides2009} for the Pokrovsky-Talapov model at finite temperatue is
based on the effective average functional renormalization group scheme by
Wetterich \cite{Wetterich1993,Berges2002}. This formalism whose equations
describe the scale dependence of the effective action has proven highly
powerful in studies of models from various fields of physics, ranging from
condensed matter theory to high-energy physics. The approach of Ref.
\cite{Lazarides2009} is based on splitting of the PT model into a sine-Gordon
part, and a part depending only on a number of solitons present. A derivation
of a functional RG transformation is relied on the assumption that the last
part is unaffected by the RG transformation. The flow equations thus obtained
reproduce well known flow equations for the sine-Gordon model \cite{Jose1977},
which belongs to the universality class of the 2D XY spin model. In this
regard, we note that the KT phase transition has been studied by the
non-perturbative RG methods on the base of a microscopic action of the
$\varphi^{4}$-model \cite{Gersdorff2001,Jakubczyk2014}, a derivative expansion
of the average action for the $O(2)$ linear $\sigma$ model \cite{Grater1995}
and for the sine-Gordon model \cite{Nagy2009}. The distinctive feature of the
nonperturbative RG techniques is that vortices are not explicitly introduced
contrary to the traditional perturbative approaches that use a mapping to the
Coulomb gas or sine-Gordon models \cite{Nagaosa}.

Regarding separation of the PT Hamiltonian in Ref. \cite{Lazarides2009} one
might note that it rules out the misfit parameter from RG transformations,
thus contradicting the previous perturbative RG studies of the model made in
the context of the CIT problem \cite{Puga1982,Horowitz1983}. Moreover, in view
of a possible application of the results to the chiral helimagnets the
splitting seems unlikely since the misfit parameter is associated with the DM
interaction, which should be taken into account in the RG analysis on an equal
footing with the exchange and Zeeman couplings. Another questionable outcome
of the separation is that the RG flow captures the KT type of behaviour, but
it remains unclear whether such a KT transition occurs in a presence of the
non-zero misfit parameter.

The current impasse lead us to reexamine the non-perturbative RG analysis of
the Pokrovsky-Talapov model. To formulate the approach based on the Wetterich
equation we exploit an equivalence of the 2D sine-Gordon and the 2D massive
Thirring (MT) models \cite{Coleman1975,Mandelstam1975}. One of the advantages
of the latter way is that the non-linear features of the sine-Gordon model now
appear in terms of a fermion interaction, what makes a sound basis for
perturbative techniques \cite{Metzner2012,Kopietz2001}, as long as the
interaction is not very strong. A key observation is that the topological term
of the PT model being linear in the scalar field variable can not be taken
into account within the Wetterich formalism which operates only quadratic or
higher order terms over fields. However, the mapping onto the Thirring
fermions converts it to the form quadratic in the Grassmann-valued fields,
whence a scaling behavior of the misfit parameter can be deduced.

Specific issues of implementation of the scheme include the following steps.
Using the results of the bosonization theory\cite{Luther1975,Witten1984} we
establish an average action corresponding to the initial PT model in the
Thirring fermion representation. Then we focus on a flow equation which
describes a scale dependence of the average action. On this stage, our
analysis is closely related to the investigation of renormalizability of the
3D Thirring model by means of the functional renormalization group formulated
in terms of the Wetterich equation \cite{Gies2010}. From the procedure we
obtain flow equations for the mass of the two-dimensional Thirring model and
the fictitious gauge field experienced by relativistic fermions, which can be
matched with the magnetic field and the strength of the DM interaction,
respectively, in the context of applications of the PT model to chiral
helimagnets. The RG transformations are complemented by flow equation for the
strength of the current-current coupling which can be compared with the
anisotropy of exchange interactions in the 2D plane.

The rigorous RG analysis is supplemented by a more physical approach based on
a duality mapping between vortices and electrostatics that was actively
exploited in the theory of the KT transition \cite{Kadanoff1978}. It is a
well-known fact that many phenomena which are difficult to interpret in the
fermion language have simple semiclassical explanations via the boson
description. It is not an exception in the present case, where we derive a
partition function of point charges, corresponding to the given PT-model of
the chiral helimagnet, and demonstrate that the DM interaction brings forth an
effective electric field directed perpendicularly to the chiral axis. A direct
consequence of occurrence of such a field is a breakdown of a KT transition
that explains an essential alternation of the RG flow pattern compared to the
sine-Gordon model.

The paper is organized as follows. In Sec. II the PT model and the
corresponding counterpart of the 2D Thirring model are formulated. Details of
the functional RG calculation are outlined in Sec.~III. In this section we
establish a picture of the RG flow using the Thirring model and the
nonperturbative RG in terms of the Wetterich equation. In Sec.~IV, we
reformulate the PT model as the model of the two-dimensional Coulomb gas and
derive RG flows in perturbative manner. Finally, we make concluding remarks in Sec.~V.

\section{Model}

The Hamiltonian of the Pokrovsky-Talapov model in notations applicable to the
two-dimensional chiral helimagnet reads as
\begin{align}
\frac{H}{T}  &  =\int d^{2}r\biggl[\frac{1}{2}\frac{J_{\perp}}{T}\left(
\partial_{x}\varphi\right)  ^{2}+\frac{1}{2}\frac{J_{||}}{T}\left(
\partial_{z}\varphi\right)  ^{2}\label{Hphi}\\
&
\,\,\,\,\,\,\,\,\,\,\,\,\,\,\,\,\,\,\,\,\,\,\,\,\,\,\,\,\,\,\,\,\,\,\,\,-\frac
{D}{T}\left(  \partial_{z}\varphi\right)  -\frac{h}{T}\cos\varphi
\biggr],\nonumber
\end{align}
where $\boldsymbol{r}=(x,z)$ and $d^{2}r=dxdz$. The first two terms correspond
to the isotropic exchange interactions within the ($x$,$z$) plane. We take
into account the anisotropy of the exchange couplings $J_{\perp}$, $J_{||}$ to
reflect related effects in thin films of the mono-axial chiral helimagnet
Cr${}_{0.33}$NbS${}_{2}$. The third term can be attributed to the DM
interaction along the $z$-axis, while the fourth describes the Zeeman energy
in a transverse field $\boldsymbol{h}=h\hat{\boldsymbol{x}}$. As it will
became clear from the RG analysis developed below the scalar field
$\varphi(\boldsymbol{r})$ should be interpreted as the polar coordinate for
the 3D unit spin vector $\boldsymbol{n}=(\cos\varphi,0,\sin\varphi)$ when
$D=0$ and for the similar vector $\boldsymbol{n}=(\cos\varphi,\sin\varphi,0)$
when the $D$ is non-zero.

Making the change $\varphi(z)=\sqrt{T/J_{\perp}}\theta(z)$, $\delta=D/T$,
$\eta=h/T$ and $\beta=\sqrt{T/J_{\perp}}$, we obtain the Euclidean action
\begin{align}
\mathcal{S}_{\text{SG}}  &  =\int d^{2}r\biggl[ \frac{1}{2}\left(
\partial_{x}\theta\right)  ^{2}+\frac{1}{2}\frac{J_{||}}{J_{\perp}}\left(
\partial_{z}\theta\right)  ^{2}\nonumber\\
&  \,\,\,\,\,\,\,\,\,\,\,\,\,\,\,\,\,\,\,\,\,\,\,\,\,\,\,\,\,\,\,\,\,\,\,\,
-\beta\delta\left(  \partial_{z}\theta\right)  -\eta\cos(\beta\theta)\biggr] .
\label{HphiS}%
\end{align}
At $J_{||}=J_{\perp}$, $\delta=0$ the 2D sine-Gordon model is restored.

The abelian bosonization rules connect 2D XY model of classical statistical
mechanics to the 2D massive Thirring model
\begin{subequations}
\begin{align}
\frac{1}{8\pi}\left(  \partial_{\mu}\theta\right)  ^{2}  &  \rightarrow
\bar{\psi}i\sigma_{\mu}\partial_{\mu}\psi,\\
\frac{1}{2\pi i}\partial_{z}\theta &  \rightarrow\bar{\psi}\sigma_{1}\psi,\\
-\eta\cos\theta &  \rightarrow im\bar{\psi}\psi,
\end{align}
where $m$ is the mass and
\end{subequations}
\begin{equation}
\sigma_{1}=\left(
\begin{array}
[c]{cc}%
0 & 1\\
1 & 0
\end{array}
\right)  ,\qquad\sigma_{2}=\left(
\begin{array}
[c]{cc}%
0 & -i\\
i & 0
\end{array}
\right)  .
\end{equation}
The Grassman valued fields are defined as
\begin{equation}
\psi=\left(
\begin{array}
[c]{c}%
\psi_{1}\\
\psi_{2}%
\end{array}
\right)  ,\qquad\bar{\psi}=\psi^{\ast}\sigma_{1}=\left(  \psi_{2}^{\ast}%
,\psi_{1}^{\ast}\right)  .
\end{equation}

After the change $\beta^{2}\mathcal{S}_{\text{SG}}/4\pi\rightarrow
\mathcal{S}_{\text{Th}}$, $\beta\theta\rightarrow\theta$, and adopting the
definitions $\left(  x_{1},x_{2}\right)  \equiv(x,z)$, $\Delta=\pi\left(
J_{||}/J_{\perp}-1\right)  $, $d=\beta^{2}\delta/2$ and $\tilde{m}=\beta
^{2}m/4\pi$ we obtain the Euclidean action of the 2D Thirring model
\begin{align}
\mathcal{S}_{\text{Th}}  &  =\int d^{2}r\biggl[\bar{\psi}\left(  i\sigma_{\mu
}D_{\mu}+i\tilde{m}\right)  \psi\nonumber\\
&  -\frac{\Delta}{2}\left(  \bar{\psi}\sigma_{1}\psi\right)  ^{2}-\frac{g}%
{2}\left(  \bar{\psi}\sigma_{\mu}\psi\right)  ^{2}\biggr]. \label{2DThirring}%
\end{align}
We call the model described by this action the modified massive Thirring model
because it contains the modified derivative $D_{\mu}=\partial_{\mu}%
-d\delta_{\mu,1}$ with the fictitious gauge field induced by the DM coupling.
The current-current interaction of the strength $g$ is also added, it involves
the conserved current $j_{\mu}=\bar{\psi}\sigma_{\mu}\psi$.

The spinor structure of the fields $\bar{\psi}$ and $\psi$ implies
\begin{equation}
\sigma_{\mu}\partial_{\mu}=\left(
\begin{array}
[c]{cc}%
0 & \partial_{1}-i\partial_{2}\\
\partial_{1}+i\partial_{2} & 0
\end{array}
\right)  .
\end{equation}
Four-fermion terms may be rewritten by using the identity
\begin{equation}
\left(  \bar{\psi}M\psi\right)  ^{2}=\text{det}M\left(  \bar{\psi}\psi\right)
^{2},
\end{equation}
where $M$ is any $2\times2$ matrix and $\text{det}M$ stands for the
determinant of $M$. \cite{Carneiro1987} This converts (\ref{2DThirring}) to
the form
\begin{align}
\mathcal{S}_{\text{Th}}  &  =\int d^{2}r\biggl[\bar{\psi}\left(  i\sigma_{\mu
}D_{\mu}+i\tilde{m}\right)  \psi\nonumber\\
&  +\left(  \frac{\Delta}{2}+g\right)  \left(  \bar{\psi}\psi\right)
^{2}\biggr]. \label{2DThirring1}%
\end{align}

By using the Fourier transforms
\begin{equation}
\psi(\boldsymbol{x})=\int\frac{d^{2}\boldsymbol{p}}{(2\pi)^{2}}%
e^{i\boldsymbol{p}\boldsymbol{x}}\psi_{\boldsymbol{p}}\equiv\int
_{p}e^{i\boldsymbol{p}\boldsymbol{x}}\psi_{\boldsymbol{p}},
\end{equation}%
\begin{equation}
\bar{\psi}(\boldsymbol{x})=\int\frac{d^{2}\boldsymbol{p}}{(2\pi)^{2}%
}e^{-i\boldsymbol{p}\boldsymbol{x}}\bar{\psi}_{\boldsymbol{p}}\equiv\int
_{p}e^{-i\boldsymbol{p}\boldsymbol{x}}\bar{\psi}_{\boldsymbol{p}},
\end{equation}
the action (\ref{2DThirring1}) takes the form in the momentum space
\begin{align}
\mathcal{S}_{\text{Th}}  &  =\int\frac{d^{2}\boldsymbol{p}}{(2\pi)^{2}}%
\bar{\psi}_{\boldsymbol{p}}\left(  -\slashed{p}-i\slashed{d}+i\tilde
{m}\right)  \psi_{\boldsymbol{p}}\nonumber\\
&  +(2\pi)^{2}\left(  \frac{\Delta}{2}+g\right)  \int\prod_{i=1}^{4}%
\frac{d^{2}\boldsymbol{p}_{i}}{(2\pi)^{2}}\left(  \bar{\psi}_{\boldsymbol{p}%
_{1}}\psi_{\boldsymbol{p}_{2}}\right)  \left(  \bar{\psi}_{\boldsymbol{p}_{3}%
}\psi_{\boldsymbol{p}_{4}}\right) \nonumber\\
&  \times\delta\left(  -\boldsymbol{p}_{1}+\boldsymbol{p}_{2}-\boldsymbol{p}%
_{3}+\boldsymbol{p}_{4}\right)  , \label{SThirring}%
\end{align}
where the slashed notations $\slashed{p}=p_{\mu}\sigma_{\mu}$ etc. for the
Dirac operators are used.

\section{Functional RG}

\subsection{Hierarchy of flow equations}

The formulation of an exact renormalization group equation is based on the
effective average action $\Gamma_{k}$, which is a generalization of the
effective action which includes only rapid modes, i.e. fluctuations with
$q^{2} \geq k^{2}$, where $k$ plays the role of an ultra-violet cutoff for
slow modes. \cite{Wetterich1993} This is achieved by adding a regulator
(infrared cutoff) $R_{k}$ to the full inverse propagator. The regulator
decouples slow modes with momenta $q^{2} \leq k^{2}$ by giving them a large
mass, while high momentum modes are not affected.

The scale dependence of $\Gamma_{k}$ is governed by the Wetterich equation
\begin{equation}
\label{WE1}\partial_{k} \Gamma_{k} = - \frac12 \text{Tr} \left[
\frac{\partial_{k} R_{k}}{\Gamma^{(2)}+R_{k}} \right]
\end{equation}
with $\Gamma^{(2)}$ indicating the second functional derivative of $\Gamma
_{k}$. The trace involves an integration over momenta as well as a summation
over internal indices. The minus sign on the right hand side of Eq.(\ref{WE1})
is due to the Grassman nature of $\bar{\psi}$ and $\psi$. \cite{Berges2002}

By definition, the average action equals the standard effective action for
$k=0$, as the infrared cutoff is absent and all fluctuations are included.
Similarly to (\ref{SThirring}) we define the effective action as
\begin{align}
\Gamma_{k}  &  =\int\frac{d^{2}\boldsymbol{p}}{(2\pi)^{2}}\bar{\psi
}_{\boldsymbol{p}}\left(  -Z_{k}\slashed{p}-i\slashed{d}_{k}+i\tilde{m}%
_{k}\right)  \psi_{\boldsymbol{p}}\nonumber\\
&  +(2\pi)^{2}\left(  \frac{\Delta_{k}}{2}+g_{k}\right)  \int\prod_{i=1}%
^{4}\frac{d^{2}\boldsymbol{p}_{i}}{(2\pi)^{2}}\left(  \bar{\psi}%
_{\boldsymbol{p}_{1}}\psi_{\boldsymbol{p}_{2}}\right)  \left(  \bar{\psi
}_{\boldsymbol{p}_{3}}\psi_{\boldsymbol{p}_{4}}\right) \nonumber\\
&  \times\delta\left(  -\boldsymbol{p}_{1}+\boldsymbol{p}_{2}-\boldsymbol{p}%
_{3}+\boldsymbol{p}_{4}\right)  , \label{Gammak}%
\end{align}
where $Z_{k}$ denotes scale dependent wave function renormalization for the
fermionic fields. All parameters in the effective action are assumed to be
scale dependent that is marked by the momentum-scale index $k$.

In practice, it is convenient to rewrite Eq.(\ref{WE1}) as
\begin{equation}
\label{Trick}\partial_{k} \Gamma_{k} = - \frac12 \tilde{\partial}_{k} \,
\text{Tr} \, \text{log} \left(  \Gamma^{(2)}+R_{k} \right)  ,
\end{equation}
where $\tilde{\partial}_{k} $ acts only on the $k$-dependence of $R_{k}$ and
not on $\Gamma^{(2)}$.

Using Eq.(\ref{Trick}) fixed points associated with the four-fermion
interactions can be simply examined. For the purpose, the inverse regularized
propagator can be split into the field-independent ($\Gamma_{k,0}^{(2)}+R_{k}%
$) and the field-dependent ($\Delta\Gamma_{k}^{(2)}$) parts. Then, the
perturbative expansion gives
\begin{align}
\frac{\partial_{k}R_{k}}{\Gamma^{(2)}+R_{k}}  &  =\tilde{\partial}%
_{k}\text{log}\left(  \Gamma_{k,0}^{(2)}+\Delta\Gamma_{k}^{(2)}+R_{k}\right)
\nonumber\\
&  =\tilde{\partial}_{k}\text{log}\left(  \Gamma_{k,0}^{(2)}+R_{k}\right)
+\tilde{\partial}_{k}\frac{\Delta\Gamma_{k}^{(2)}}{\Gamma_{k,0}^{(2)}+R_{k}%
}\nonumber\\
&  -\frac{1}{2}\tilde{\partial}_{k}\left(  \frac{\Delta\Gamma_{k}^{(2)}%
}{\Gamma_{k,0}^{(2)}+R_{k}}\right)  ^{2}+\ldots\label{Series}%
\end{align}

In the computation of the RG flow equations a regulator function needs to be
specified which determines the regularization scheme. For the relativistic
fermions we may choose \cite{Gies2010,Braun2012}
\begin{equation}
R_{k}=-\frac{\delta_{p,q}}{(2\pi)^{2}}Z_{k}r\left(  \frac{q^{2}}{k^{2}%
}\right)  \left(
\begin{array}
[c]{cc}%
0 & \slashed{q}^{T}\\
\slashed{q} & 0
\end{array}
\right)  .
\end{equation}
From the explicit calculations given in Appendix \ref{AppendixA} we find the
propagator
\begin{align}
&  \left(  \Gamma_{k,0}^{(2)}+R_{k}\right)  ^{-1}\nonumber\\
&  =\left(  2\pi\right)  ^{2}\delta_{p,q}\left(
\begin{array}
[c]{cc}%
0 & \frac{\slashed{\alpha}_{k}-i\tilde{m}_{k}}{\alpha_{k}^{2}+\tilde{m}%
_{k}^{2}}\\
\frac{\slashed{\beta}_{k}^{T}+i\tilde{m}_{k}}{\beta_{k}^{2}+\tilde{m}_{k}^{2}}
& 0
\end{array}
\right) \nonumber\\
&  \equiv\left(  2\pi\right)  ^{2}\hat{G}_{0}\delta_{p,q},\label{InverseM}
\end{align}
where
\begin{align*}
\slashed{\alpha}_{k}  &  =-Z_{k}\slashed{q}\left[  1+r\left(  \frac{q^{2}%
}{k^{2}}\right)  \right]  -i\slashed{d}_{k}\\
&  \equiv a_{k}\slashed{q}-i\slashed{d}_{k},
\end{align*}%
\begin{align*}
\slashed{\beta}_{k}  &  =-Z_{k}\slashed{q}\left[  1+r\left(  \frac{q^{2}%
}{k^{2}}\right)  \right]  +i\slashed{d}_{k}\\
&  \equiv a_{k}\slashed{q}+i\slashed{d}_{k},
\end{align*}
and $\alpha^{2}=\alpha_{1}^{2}+\alpha_{2}^{2}$, $\beta^{2}=\beta_{1}^{2}%
+\beta_{2}^{2}$.

Similar derivation of the field-dependent part yields
\begin{align}
\Delta\Gamma_{k}^{(2)}  &  =2(2\pi)^{2}\left(  \frac{\Delta_{k}}{2}%
+g_{k}\right) \nonumber\\
&  \times\left(
\begin{array}
[c]{cc}%
-\bar{\psi}^{T}\bar{\psi} & \bar{\psi}^{T}\psi^{T}+\psi^{T}\bar{\psi}^{T}\\
\psi\bar{\psi}+\bar{\psi}\psi & -\psi\psi^{T}%
\end{array}
\right)  \delta_{p,q}\nonumber\\
&  =2(2\pi)^{2}\left(  \frac{\Delta_{k}}{2}+g_{k}\right)  \hat{G}_{1}%
\delta_{p,q} \label{FDP}%
\end{align}
where the second functional derivative is evaluated for homogeneous (constant)
background fields. In momentum space it means that $\Delta\Gamma_{k}^{(2)}$ is
evaluated at
\begin{equation}
\psi_{\boldsymbol{p}}=(2\pi)^{2}\psi\delta(\boldsymbol{p}),\qquad\bar{\psi
}_{\boldsymbol{p}}=(2\pi)^{2}\bar{\psi}\delta(\boldsymbol{p}),
\label{GrasConst}%
\end{equation}
where $\psi$ and $\bar{\psi}$ on the right-hand side are constant.
\cite{Braun2012}

We can then expand the flow equation in powers of the Grassman fields by
combining Eqs.(\ref{Trick},\ref{Series})
\begin{align}
\partial_{k}\Gamma_{k}  &  =-\frac{1}{2}\text{Tr}\left[  \tilde{\partial}%
_{k}\text{log}\left(  \Gamma_{k,0}^{(2)}+R_{k}\right)  \right]  -\frac{1}%
{2}\text{Tr}\left[  \tilde{\partial}_{k}\frac{\Delta\Gamma_{k}^{(2)}}%
{\Gamma_{k,0}^{(2)}+R_{k}}\right] \nonumber\\
&  +\frac{1}{4}\text{Tr}\left[  \tilde{\partial}_{k}\left(  \frac{\Delta
\Gamma_{k}^{(2)}}{\Gamma_{k,0}^{(2)}+R_{k}}\right)  ^{2}\right]  +\ldots.
\label{PT1}%
\end{align}
The powers of $\frac{\Delta\Gamma_{k}^{(2)}}{\Gamma_{k,0}^{(2)}+R_{k}}$ can be
calculated by simple matrix multiplications. The RG flow equations can now be
computed straightforwardly by comparing the coefficients of the fermion
interaction terms of the right-hand side of Eq. (\ref{PT1}) with the couplings
included in the ansatz (\ref{Gammak}).

\subsection{Two-fermion beta functions}

Let us now derive the RG flow equations for the couplings that involved in the
part of the action which is quadratic in the fermionic fields $\psi$ and
$\bar{\psi}$. From the series (\ref{PT1}) it is clear that only the term
\begin{equation}
-\frac{1}{2}\text{Tr}\left[  \frac{\Delta\Gamma_{k}^{(2)}}{\Gamma_{k,0}%
^{(2)}+R_{k}}\right]  =-\left(  \frac{\Delta_{k}}{2}+g_{k}\right)  \Omega
\int\frac{d^{2}q}{(2\pi)^{2}}\text{Tr}\left[  \hat{G}_{1}\hat{G}_{0}\right]  ,
\label{PT1st}%
\end{equation}
where $\Omega$ is the volume of the system, contributes to the RG flow of the
needed couplings, and $\hat{G}_{1}$ is defined by Eq. (\ref{FDP}).

An elementary calculation gives
\begin{align*}
\text{Tr}\left[  \hat{G}_{1}\hat{G}_{0}\right]   &  =-\frac{\alpha_{k\mu}%
}{\alpha_{k}^{2}+\tilde{m}_{k}^{2}}\left(  \bar{\psi}\sigma_{\mu}\psi\right)
+\frac{\beta_{k\mu}}{\beta_{k}^{2}+\tilde{m}_{k}^{2}}\left(  \bar{\psi}%
\sigma_{\mu}\psi\right) \\
&  -i\tilde{m}_{k}\left(  \frac{1}{\alpha_{k}^{2}+\tilde{m}_{k}^{2}}+\frac
{1}{\beta_{k}^{2}+\tilde{m}_{k}^{2}}+\right)  \bar{\psi}\psi
\end{align*}%
\begin{align*}
&  =-2i\tilde{m}_{k}\frac{a_{k}^{2}q^{2}-d_{k}^{2}+\tilde{m}_{k}^{2}}{\left(
a_{k}^{2}q^{2}-d_{k}^{2}+\tilde{m}_{k}^{2}\right)  ^{2}+4a_{k}^{2}d_{k}%
^{2}q_{1}^{2}}\left(  \bar{\psi}\psi\right) \\
&  +2id_{k}\frac{a_{k}^{2}q^{2}-d_{k}^{2}+\tilde{m}_{k}^{2}-2a_{k}^{2}%
q_{1}^{2}}{\left(  a_{k}^{2}q^{2}-d_{k}^{2}+\tilde{m}_{k}^{2}\right)
^{2}+4a_{k}^{2}d_{k}^{2}q_{1}^{2}}\left(  \bar{\psi}\sigma_{1}\psi\right)
\end{align*}%
\begin{equation}
-4id_{k}\frac{a_{k}^{2}q_{1}q_{2}}{\left(  a_{k}^{2}q^{2}-d_{k}^{2}+\tilde
{m}_{k}^{2}\right)  ^{2}+4a_{k}^{2}d_{k}^{2}q_{1}^{2}}\left(  \bar{\psi}%
\sigma_{2}\psi\right)  .
\end{equation}
The third term drops out of Eq.(\ref{PT1st}) after integration over the
momentum $\boldsymbol{q}$ that brings forth
\begin{align}
&  -\frac{1}{2}\text{Tr}\left[  \tilde{\partial}_{k}\frac{\Delta\Gamma
_{k}^{(2)}}{\Gamma_{k,0}^{(2)}+R_{k}}\right] \nonumber\\
&  =2i\Omega\tilde{m}_{k}\left(  \frac{\Delta_{k}}{2}+g_{k}\right)
\tilde{\partial}_{k}\mathcal{L}_{1}\left(  \bar{\psi}\psi\right) \nonumber\\
&  -2i\Omega d_{k}\left(  \frac{\Delta_{k}}{2}+g_{k}\right)  \tilde{\partial
}_{k}\mathcal{L}_{2}\left(  \bar{\psi}\sigma_{1}\psi\right)  , \label{RHS1}%
\end{align}
where the threshold functions are
\begin{equation}
\mathcal{L}_{1}=\int\frac{d^{2}q}{(2\pi)^{2}}\frac{a_{k}^{2}q^{2}-d_{k}%
^{2}+\tilde{m}_{k}^{2}}{\left(  a_{k}^{2}q^{2}-d_{k}^{2}+\tilde{m}_{k}%
^{2}\right)  ^{2}+4a_{k}^{2}d_{k}^{2}q_{1}^{2}}, \label{TFL1}%
\end{equation}%
\begin{equation}
\mathcal{L}_{2}=\int\frac{d^{2}q}{(2\pi)^{2}}\frac{a_{k}^{2}q^{2}-d_{k}%
^{2}+\tilde{m}_{k}^{2}-2a_{k}^{2}q_{1}^{2}}{\left(  a_{k}^{2}q^{2}-d_{k}%
^{2}+\tilde{m}_{k}^{2}\right)  ^{2}+4a_{k}^{2}d_{k}^{2}q_{1}^{2}}.
\label{TFL2}%
\end{equation}

The ansatz for the kinetic term in the effective action (\ref{Gammak}) gives
\begin{equation}
\label{LHS1}\partial_{k} \Gamma_{k} = -i \Omega\left(  \partial_{k} d_{k}
\right)  \left(  \bar{\psi} \sigma_{1} \psi\right)  + i \Omega\left(
\partial_{k} \tilde{m}_{k} \right)  \bar{\psi} \psi.
\end{equation}
In our approximation, the RG running of $Z$ is trivial, i.e. $\partial_{k} Z
=0$. Thus, the associated anomalous dimension, $\eta= -k \partial_{k} \ln Z$,
is zero. Therefore, in what follows, we set the wave-function renormalization
to one, $Z=1$.

Comparing coefficients of the quadratic contributions (\ref{RHS1}, \ref{LHS1})
to the exact flow equations yields
\begin{equation}
\label{RGm}\partial_{k} m_{k} = 2 \left(  \frac{\Delta_{k}}{2} + g_{k}
\right)  m_{k} \tilde{\partial}_{k} \mathcal{L}_{1},
\end{equation}
\begin{equation}
\label{RGd}\partial_{k} d_{k} = 2 \left(  \frac{\Delta_{k}}{2} + g_{k}
\right)  d_{k} \tilde{\partial}_{k} \mathcal{L}_{2}.
\end{equation}

\subsection{Four-fermion beta function}

Formula for the four-fermion beta function reads
\begin{align}
&  \frac{1}{4}\text{Tr}\left[  \tilde{\partial}_{k}\left(  \frac{\Delta
\Gamma_{k}^{(2)}}{\Gamma_{k,0}^{(2)}+R_{k}}\right)  ^{2}\right] \nonumber\\
&  =\left(  \frac{\Delta_{k}}{2}+g_{k}\right)  ^{2}\Omega\int\frac{d^{2}%
q}{(2\pi)^{2}}\text{Tr}\left[  \hat{G}_{1}\hat{G}_{0}\hat{G}_{1}\hat{G}%
_{0}\right]  . \label{PT2}%
\end{align}
By evaluating
\begin{align}
&  \text{Tr}\left[  \hat{G}_{1}\hat{G}_{0}\hat{G}_{1}\hat{G}_{0}\right]
\nonumber\\
&  =\biggl[-\text{det}\left(  \frac{\slashed{\alpha}_{k}-i\tilde{m}_{k}%
}{\alpha_{k}^{2}+\tilde{m}_{k}^{2}}\right) \nonumber\\
&  -\text{det}\left(  \frac{\slashed{\beta}_{k}-i\tilde{m}_{k}}{\beta_{k}%
^{2}+\tilde{m}_{k}^{2}}\right) \nonumber\\
&  +2\mathcal{X}\left(  \frac{\slashed{\alpha}_{k}-i\tilde{m}_{k}}{\alpha
_{k}^{2}+\tilde{m}_{k}^{2}},\frac{\slashed{\beta}_{k}+i\tilde{m}_{k}}%
{\beta_{k}^{2}+\tilde{m}_{k}^{2}}\right)  \biggr]\left(  \bar{\psi}%
\psi\right)  ^{2},
\end{align}
where the function with the matrix arguments is defined by
\begin{align}
\mathcal{X}\left(  \hat{M},\hat{N}\right)   &  =\frac{1}{2}(M_{11}%
N_{22}-M_{12}N_{21}\nonumber\\
&  -M_{21}N_{12}+M_{22}N_{11}),
\end{align}
the four-fermion terms are straightforwardly appear.

By employing the ansatz (\ref{Gammak}) for the effective action after some
elementary algebra we get for the constant fields (\ref{GrasConst})
\begin{equation}
\label{LHSPT2}\Gamma_{k} = \Omega\left(  \frac{\Delta_{k}}{2} + g_{k} \right)
\left(  \bar{\psi} \psi\right)  ^{2}.
\end{equation}

The flow for the coupling constant $\lambda_{k}=\left(  {\Delta_{k}}/{2}%
+g_{k}\right)  $ is obtained by comparing both sides of Eq.(\ref{PT1}) and by
using Eqs.(\ref{PT2},\ref{LHSPT2})
\begin{equation}
\partial_{k}\lambda_{k}=\lambda_{k}^{2}\tilde{\partial}_{k}\mathcal{L}_{3}.
\label{RGlambda}%
\end{equation}
The flow involves the threshold function
\begin{align}
\mathcal{L}_{3}  &  =\int\frac{d^{2}q}{(2\pi)^{2}}\biggl[-\text{det}\left(
\frac{\slashed{\alpha}_{k}-i\tilde{m}_{k}}{\alpha_{k}^{2}+\tilde{m}_{k}^{2}%
}\right) \nonumber\\
&  -\text{det}\left(  \frac{\slashed{\beta}_{k}-i\tilde{m}_{k}}{\beta_{k}%
^{2}+\tilde{m}_{k}^{2}}\right) \nonumber\\
&  +2\mathcal{X}\left(  \frac{\slashed{\alpha}_{k}-i\tilde{m}_{k}}{\alpha
_{k}^{2}+\tilde{m}_{k}^{2}},\frac{\slashed{\beta}_{k}+i\tilde{m}_{k}}%
{\beta_{k}^{2}+\tilde{m}_{k}^{2}}\right)  \biggr]. \label{TFL3}%
\end{align}

Eqs. (\ref{RGm}, \ref{RGd}, \ref{RGlambda}) represent the main result of the
two sections.

\subsection{Solution of the flow equations}

For practical computations of the flow equations we use the sharp cutoff
regulator
\begin{equation}
\label{Sharp}r \left(  \frac{q^{2}}{k^{2}} \right)  = \left\{
\begin{array}
[c]{cc}%
\infty, & q^{2} < k^{2}\\
0, & q^{2} > k^{2}%
\end{array}
\right.  ,
\end{equation}
which facilitates explicit evaluation of the threshold functions
$\mathcal{L}_{1,2,3}$, which encode the details of the regularization scheme.
Their detailed derivations are given in the Appendix \ref{AppendixB}.

To solve numerically the flow equations and look for possible fixed points, it
is convenient to introduce the dimensionless quantities (see Appendix
\ref{AppendixC}), where all quantities are expressed in units of the running
scale $k$,
\begin{equation}
\bar{\lambda}=\lambda_{k},\qquad\bar{m}=k^{-1}\tilde{m}_{k},\qquad\bar
{d}=k^{-1}d_{k}.
\end{equation}
In terms of these variables the flow equations write
\begin{equation}
\partial_{t}\bar{m}=\bar{m}+\frac{\bar{m}\bar{\lambda}}{2\pi\bar{d}^{2}}%
\frac{\left(  1-\bar{d}^{2}+\bar{m}^{2}\right)  }{\sqrt{\left[  2\left(
\frac{1-\bar{d}^{2}+\bar{m}^{2}}{2\bar{d}}\right)  ^{2}+1\right]  ^{2}-1}},
\label{WRG1}%
\end{equation}%
\begin{align}
\partial_{t}\bar{d}  &  =\bar{d}-\frac{\bar{\lambda}}{2\pi\bar{d}}\nonumber\\
&  +\frac{\bar{\lambda}}{2\pi\bar{d}}\frac{\left(  1-\bar{d}^{2}+\bar{m}%
^{2}\right)  \left(  1+\frac{1}{2\bar{d}^{2}}\left[  1-\bar{d}^{2}+\bar{m}%
^{2}\right]  \right)  }{\sqrt{\left[  2\left(  \frac{1-\bar{d}^{2}+\bar{m}%
^{2}}{2\bar{d}}\right)  ^{2}+1\right]  ^{2}-1}}, \label{WRG2}%
\end{align}%
\begin{equation}
\partial_{t}\bar{\lambda}=-\frac{\bar{\lambda}^{2}}{\pi}\left(  1-\frac
{\bar{m}^{2}}{\bar{d}^{2}}\right)  \frac{1}{\sqrt{\left[  2\left(
\frac{1-\bar{d}^{2}+\bar{m}^{2}}{2\bar{d}}\right)  ^{2}+1\right]  ^{2}-1}},
\label{WRG3}%
\end{equation}
where $|t|=\ln\left(  \Lambda/k\right)  $

Before discussing the results for the model (\ref{2DThirring1}) we first focus
on a more elementary functional RG flow for the massive Thirring (or
equivalently, the sine-Gordon) model. First we realize that if $\bar{d}=0$ the
RG trajectories remain in the plane ($\bar{m}$, $\bar{\lambda}$). Indeed, in
this case the flow equations turn into
\begin{equation}
\label{KT1}\partial_{t} \bar{m} = \bar{m} \left[  1+ \frac{\bar{\lambda}}%
{\pi(1+\bar{m}^{2})} \right]  ,
\end{equation}
\begin{equation}
\label{KT2}\partial_{t} \bar{\lambda} = \frac{2}{\pi} \frac{\bar{\lambda}^{2}
\bar{m}^{2}}{\left(  1 + \bar{m}^{2} \right)  ^{2}}.
\end{equation}

At this point, we make the important observation that the beta functions in
Eqs.(\ref{WRG1},\ref{WRG2},\ref{WRG3}) have \textit{the multiplicative
regulator dependence} (see also a discussion in Ref. \cite{Gies2010}). In
addition, we note that the sine-Gordon and the massive Thirring models are
equivalent provided the coupling constants $\beta$ and $g$ of the two models
are related through the relation \cite{Coleman1975,Mandelstam1975}
\begin{equation}
\frac{4\pi}{\beta^{2}} = 1 +\frac{g}{\pi}.
\end{equation}
As for the sine-Gordon model, the system undergoes a Kosterlitz-Thouless
continuous phase transition at $\beta^{2} = 8\pi$. Given the equivalence
between the SG and MT models, the transition point for the massive Thirring
model takes place at $g=-\pi/2$. \cite{Yoshida2002} For $\beta^{2} < 8\pi$
($g>-\pi/2$) the coupling constant $g$ flows to strong coupling that indicates
the opening of a gap in the spectrum, and the relevant degrees of freedom are
massive fermionic solitons. For $\beta^{2} > 8\pi$ ($g<-\pi/2$) the
weak-coupling regime arises, where the coupling $g$ flows to zero, and the
relevant degrees of freeedom are massless bosons.

\begin{figure}[t]
\begin{center}
\includegraphics[width=70mm]{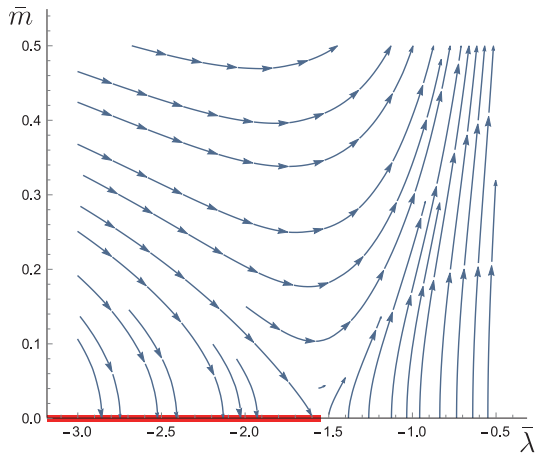}
\end{center}
\caption{ Flow diagrams for the RG Eqs. \ref{SKT1} and \ref{SKT2}.\quad\ The
red line indicates a line of fixed-points. }%
\label{Fig0}%
\end{figure}

Taking heed of these features, we apply the another rescaling in
Eqs.(\ref{WRG1},\ref{WRG2},\ref{WRG3}),
\begin{equation}
\bar{\lambda}\rightarrow2\bar{\lambda},\qquad\bar{m}\rightarrow2\bar{m}%
,\qquad\bar{d}\rightarrow2\bar{d},
\end{equation}
to eliminate the multiplicative regulator dependence and reach a consistency
between the SG and MT theories. This modifies Eqs.(\ref{KT1},\ref{KT2}) to
\begin{equation}
\partial_{t}\bar{m}=\bar{m}\left[  1+\frac{2\bar{\lambda}}{\pi(1+4\bar{m}%
^{2})}\right]  , \label{SKT1}%
\end{equation}%
\begin{equation}
\partial_{t}\bar{\lambda}=\frac{16}{\pi}\frac{\bar{\lambda}^{2}\bar{m}^{2}%
}{\left(  1+4\bar{m}^{2}\right)  ^{2}}. \label{SKT2}%
\end{equation}
These flow equations reproduce well-known scaling equations of the KT type.
The corresponding flow diagram is shown in Fig. \ref{Fig0}, where there exists
a line of fixed points with $\bar{m}=0$ and finite $\bar{\lambda}^{\ast}%
<\bar{\lambda}_{\text{KT}}$, where $\bar{\lambda}_{\text{KT}}=-\pi/2$.

\begin{figure}[t]
\begin{center}
\includegraphics[width=90mm]{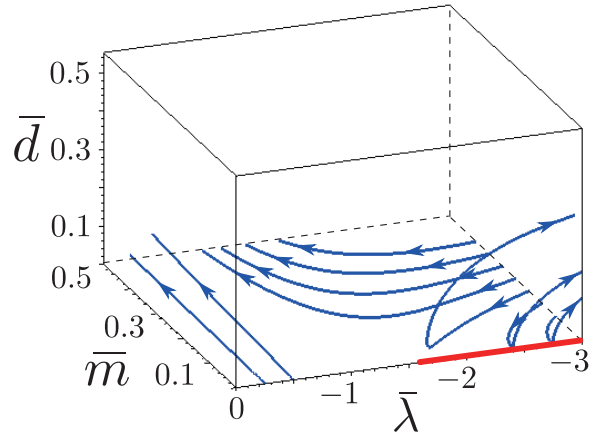}
\end{center}
\caption{The RG trajectories of the 2D Thirring-like PT model. The red line
indicates a line of fixed-points in the case where the linear gradient term is
absent. It is seen that the flows ultimately escape from the line of
fixed-points toward the high-temperature phase.}%
\label{RgFRG}%
\end{figure}

In order to determine the fate of the Kosterlitz-Thouless transition in the
presence of the linear gradient term we make use of all the RG equations
\begin{equation}
\partial_{t}\bar{m}=\bar{m}+\frac{\bar{m}\bar{\lambda}}{4\pi\bar{d}^{2}}%
\frac{\left(  1-4\bar{d}^{2}+4\bar{m}^{2}\right)  }{\sqrt{\left[  2\left(
\frac{1-4\bar{d}^{2}+4\bar{m}^{2}}{4\bar{d}}\right)  ^{2}+1\right]  ^{2}-1}},
\label{SWRG1}%
\end{equation}%
\begin{align}
\partial_{t}\bar{d}  &  =\bar{d}-\frac{\bar{\lambda}}{4\pi\bar{d}}\nonumber\\
&  +\frac{\bar{\lambda}}{4\pi\bar{d}}\frac{\left(  1-4\bar{d}^{2}+4\bar{m}%
^{2}\right)  \left(  1+\frac{1}{8\bar{d}^{2}}\left[  1-4\bar{d}^{2}+4\bar
{m}^{2}\right]  \right)  }{\sqrt{\left[  2\left(  \frac{1-4\bar{d}^{2}%
+4\bar{m}^{2}}{4\bar{d}}\right)  ^{2}+1\right]  ^{2}-1}}, \label{SWRG2}%
\end{align}%
\begin{equation}
\partial_{t}\bar{\lambda}=-\frac{2\bar{\lambda}^{2}}{\pi}\left(  1-\frac
{\bar{m}^{2}}{\bar{d}^{2}}\right)  \frac{1}{\sqrt{\left[  2\left(
\frac{1-4\bar{d}^{2}+4\bar{m}^{2}}{4\bar{d}}\right)  ^{2}+1\right]  ^{2}-1}}.
\label{SWRG3}%
\end{equation}
The corresponding parametric flow is shown in Fig. \ref{RgFRG}. At finite
$\bar{d}$ value, the flow is seen to initially closely follow the KT flow at
$\bar{d}=0$, approach the fixed line at $\bar{m}=0$, but ultimately escape
from the line toward the high-temperature phase. We may conclude that in a
presence of the linear gradient term (DM interaction) no KT transition can exist.

\section{Coulomb gas model}

A remarkable feature found in early studies
\cite{Berezinskii1971,Kosterlitz1973,Kosterlitz1974,Jose1977} is that the
defect-mediated transition of the 2D XY model and its analogues can be mapped
to the insulator-conductor transition of a two-dimensional Coulomb gas. To
elucidate the origin of the flow of the Thirring model we formulate the 2D
Coulomb gas model by using discrete vector calculus on a square lattice for
the Hamiltonian (\ref{2DThirring1}), where, for simplicity, we restrict the
representation by the isotropic case, $J_{||}=J_{\perp}$ . For definitness,
all sums run over the sites of a square lattice although the transformation
described below are easily generalized.

The partition function defined on such a lattice is of the form
\begin{align}
\mathcal{Z}  &  =\int\mathcal{D}\varphi\,\exp\biggl[\beta\tilde{J}%
\sum_{\langle ij\rangle}\cos\left(  \varphi_{i}-\varphi_{j}-\alpha_{ij}\right)
\nonumber\\
&  \;\;\;\;+\beta h\sum_{i}\cos\varphi_{i}\biggr], \label{PHIMOD}%
\end{align}
where henceforth $\beta=(k_{\text{B}}T)^{-1}$ is the inverse temperature,
$\tilde{J}=S^{2}\sqrt{J^{2}+D^{2}}$ is the effective exchange parameter,
$h=Sh_{x}$, and the bond angle is given by $\alpha_{ij}=\tan^{-1}\left(
D/J\right)  $ for the $ij$-link along the $z$-axis and zero otherwise. The
first sum runs over all nearest neighbor sites within the ($xz$)-plane.

A duality mapping onto the Coulomb gas model is derived in detail in Appendix
\ref{AppendixD}. The resulting partition function for point charge particles
reads as
\begin{align}
\mathcal{Z}_{\text{eff}}  &  =\sum_{\left\{  q(\boldsymbol{r})\right\}
}\,\exp\biggl[\pi K_{0}\sum_{\boldsymbol{r}\not =\boldsymbol{r}^{\prime}%
}q(\boldsymbol{r})\log\left\vert \boldsymbol{r}-\boldsymbol{r}^{\prime
}\right\vert q(\boldsymbol{r}^{\prime})\nonumber\\
&  \;\;\;\;+2\pi\sum_{\boldsymbol{r}}(\boldsymbol{E}_{x}\cdot\boldsymbol{r}%
)q(\boldsymbol{r})\biggr]y_{0}^{\sum_{\boldsymbol{r}}\,q^{2}(\boldsymbol{r})},
\label{QGPF}%
\end{align}
where the bare values for the vortex coupling $K_{0}=\beta\tilde{J}$ and for
the vortex pair fugacity $y_{0}=\exp\left(  -\beta\pi^{2}\tilde{J}/2\right)  $
. The first term in the exponential of Eq.(\ref{QGPF}) describes the
charge-charge interactions, the second includes the sum of the self-energies
associated with the each elementary charge $q(\boldsymbol{r})$, which arises
due to the effective uniform $x$-direction field, $\boldsymbol{E}_{x}%
=\alpha\beta\tilde{J}\boldsymbol{e}_{x}$. The question whether a topological
order realizes is thereby mapped, as in the conventional Kosterlitz-Thouless
transition, to the problem of screening in the Coulomb gas, albeit now with
modified terms due to the DM interaction.

The scaling equations can then be obtained from (\ref{QGPF}) in the limit of
small fugacity. The procedure parallels what was detailed in Refs.
\cite{Sujani1994,Atland2006}. At a general minimum scale $a=e^{l}$ the
renormalized vortex coupling $K_{l}$, the vortex pair fugacity $y_{l}$ and the
topological electric field, $\epsilon_{l}$, obey the scaling equations
\begin{align}
\frac{dK_{l}}{dl}  &  =-4\pi^{3}K_{l}^{2}y_{l}^{2},\\
\frac{dy_{l}}{dl}  &  =\left(  2-\pi K_{l}+\pi\epsilon_{l}\right)  y_{l},\\
\frac{d\epsilon_{l}}{dl}  &  =\epsilon_{l}+4\pi^{3}y_{l}^{2}K_{l}\epsilon_{l}.
\end{align}

\begin{figure}[t]
\begin{center}
\includegraphics[width=90mm]{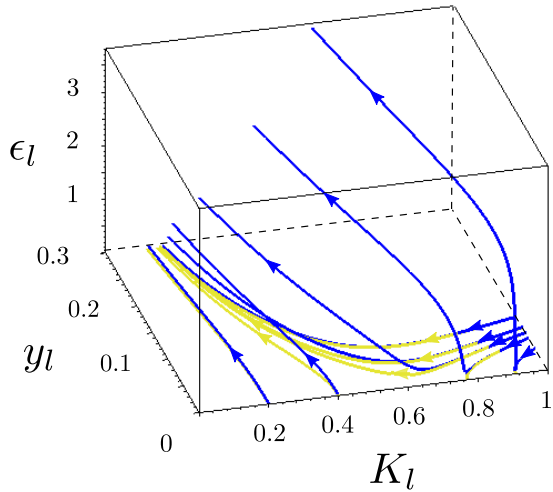}
\end{center}
\caption{The RG flows of the 2D Coulomb gas counterpart of the PT model (blue)
and for the XY model (yellow), $\epsilon_{l}=0$. We see that the RG flows
quickly escape from the line of fixed points, which is indicated by a red
line.}%
\label{RgCG}%
\end{figure}

Fig. \ref{RgCG}. shows numerical solutions of RG flows. For $\epsilon=0$ it
reproduces the well-known $K_{l}(y_{l})$ scale dependence of the KT-theory
that tend asymptotically to zero (infinity) for $T>T_{\text{KT}}$,
$k_{\text{B}} T_{\text{KT}}=\pi J/2$, and to constant (zero) for
$T<T_{\text{KT}}$. The results for non-zero electric fields are also presented
in this Figure. Apparently, at all temperatures $K_{l}$ tends asymptotically
to zero and $y_{l}$ tends to infinity, i.e. vortex pairs are unbound by the
electric field. It means that the breaking apart of dipoles by the topological
field begins to exceed the vortex attractive interaction. In addition, the
plot clearly demonstrates that the both models, the Thirring model
(\ref{2DThirring1}) with a fictitious gauge field and the Coulomb gas in an
electric field, are in the same universality class.

This concludes the RG analysis which shows that the DM interaction is
relevant, it creates an effective electric field perpendicular to the
direction of the DM vector on a lattice and eliminates the KT transition.

\section{Concluding remarks}

In this paper we investigated the effects of the linear gradient term on a
topological Kosterlitz-Thouless transition based on the functional RG approach
by Wetterich. Our work has been motivated by some experimental studies on the
topological defects or vortex excitations which may occur in thin films of
chiral helimagnets \cite{Leonov2016,Fukui2016,Kato2017}. This class of
phenomena are deeply related to the Pokrovsky-Talapov model.

Our main result is that a nonzero linear gradient term (or misfit parameter)
of the PT-model, which can be related with the Dzyaloshinsky-Moriya
interaction in magnetis systems, makes such a transition prohibited, what
contradicts the previous consideration of this problem by the non-perturbative
RG methods.

In order to argue this conclusion the initial PT model has been reformulated
in terms of the 2D theory of relativistic fermions using an analogy between
the 2D sine-Gordon and the massive Thirring models. In the new formalism the
misfit parameter corresponds to an effective gauge field that enables to
include it in the renormalization-group procedure on an equal footing with the
other parameters of the theory. With the new fermionic action in hands, we
apply the Wetterich equation to obtain flow equations for the parameters of
the action. We demonstrate that these RG equations reproduce a KT type of
behavior if the misfit equals to zero. However, any small nonzero value of the
quantity rules out a possibility of the KT transition. To confirm the finding
we develop a description of the issue in terms of the 2D Coulomb gas model
using the corresponding mapping for the Pokrovsky-Talapov model. Within the
approach the breakdown of the KT scenario gains a transparent meaning. The
misfit parameter results in an appearance of an effective electric field lying
in the plane what prevents a formation of bound vortex-antivortex dipoles. It
is worth noting an advantage of the non-perturbative RG approach in comparison
to the RG of the model of the 2D Coulomb gas. The former does not require a
smallness of the magnetic field.

In closing, a matter of interest is a further application of a functional RG
technique to study spin models describing a behavior of real chiral magnetic
systems. In this regard we note the recent work \cite{Hering2017}, where the
functional renormalization group analysis of Dzyaloshinsky-Moriya and
Heisenberg spin interactions on the kagome lattice has been done.

\begin{acknowledgments}
Special thanks are due to Y. Togawa for informative discussions in early stage
of the works. The work was supported by the Government of the Russian
Federation Program 02.A03.21.0006. This work was also supported by JSPS and
RFBR under the Japan-Russia Research Cooperative Program, and JSPS
Core-to-Core Program, A. Advanced Research Networks. This work was supported
by JSPS KAKENHI Grant Numbers JP17H02923 and JP25220803. I. P. acknowledges
financial support by Ministry of Education and Science of the Russian
Federation, Grant No. 316 MK-6230.2016.2. A.S.O. and P.A.N. acknowledge
funding by the RFBR, Grant 17-52-50013.
\end{acknowledgments}

\appendix

\section{The inverse propagator}

\label{AppendixA}

The matrix of second derivatives of $\Gamma_{k}$ with respect to the fermion
fields deduced from Eq.(\ref{Gammak})
\begin{equation}
\Gamma_{k}^{(2)}(p,q)=\left(
\begin{array}
[c]{cc}%
\overrightarrow{\partial}_{\psi_{-p}^{T}}\Gamma_{k}\overleftarrow{\partial
}_{\psi_{q}} & \overrightarrow{\partial}_{\psi_{-p}^{T}}\Gamma_{k}%
\overleftarrow{\partial}_{\bar{\psi}_{-q}^{T}}\\
\overrightarrow{\partial}_{\bar{\psi}_{p}}\Gamma_{k}\overleftarrow{\partial
}_{\psi_{q}} & \overrightarrow{\partial}_{\bar{\psi}_{p}}\Gamma_{k}%
\overleftarrow{\partial}_{\bar{\psi}_{-q}^{T}}%
\end{array}
\right)
\end{equation}
results in the field-independent part
\begin{gather}
\Gamma_{k,0}^{(2)}(p,q)=\frac{\delta_{p,q}}{(2\pi)^{2}}\nonumber\\
\times\left(
\begin{array}
[c]{cc}%
0 &
\begin{array}
[c]{c}%
-Z_{k}\slashed{p}^{T}\\
+i\slashed{d}^{T}-i\tilde{m}%
\end{array}
\\%
\begin{array}
[c]{c}%
-Z_{k}\slashed{p}\\
-i\slashed{d}+i\tilde{m}%
\end{array}
& 0
\end{array}
\right)  .
\end{gather}
Then the inverted form of the regularized propagator reads {
\begin{gather}
\left(  \Gamma_{k,0}^{(2)}+R_{k}\right)  (p,q)=\frac{\delta_{p,q}}{(2\pi)^{2}%
}\nonumber\\
\times\left(
\begin{array}
[c]{cc}%
0 &
\begin{array}
[c]{c}%
-Z_{k}\slashed{p}^{T}\left[  1+r\left(  \frac{q^{2}}{k^{2}}\right)  \right] \\
+i\slashed{d}^{T}-i\tilde{m}%
\end{array}
\\%
\begin{array}
[c]{c}%
-Z_{k}\slashed{p}\left[  1+r\left(  \frac{q^{2}}{k^{2}}\right)  \right] \\
-i\slashed{d}+i\tilde{m}%
\end{array}
& 0
\end{array}
\right)  ,
\end{gather}
} the inverse of the matrix yields the result (\ref{InverseM}). To find the
form of the field-dependent part (\ref{FDP}) the property $\bar{\psi}%
\psi=-\psi^{T}\bar{\psi}^{T}$ appears to be useful.

\section{Treshold functions}

\label{AppendixB}

The flow equations include single integrals due to one-loop structure of
Wetterich equation, the threshold functions, which contain details of the
regularization scheme. The definitions of the threshold functions are given by
Eqs.(\ref{TFL1},\ref{TFL2},\ref{TFL3}). In the flow equations $\tilde
{\partial}_{k}$ is defined to act on the regulator's $k$-dependence. The sharp
cut-off regulator (\ref{Sharp}) has the remarkable feature that all threshold
integrals can be done explicitly.

Indeed, in the polar coordinates
\begin{align}
&  \tilde{\partial}_{k}\mathcal{L}_{1}\nonumber\\
&  =\tilde{\partial}_{k}\int_{k}^{\Lambda}\frac{dq}{(2\pi)^{2}}\int_{0}^{2\pi
}d\phi\frac{q\left(  q^{2}-d_{k}^{2}+\tilde{m}_{k}^{2}\right)  }{\left(
q^{2}-d_{k}^{2}+\tilde{m}_{k}^{2}\right)  ^{2}+4d_{k}^{2}q^{2}\cos^{2}\phi},
\end{align}
where $\Lambda$ is the ultraviolet cutoff, and we take into account that
$a=-1$ for the regulator (\ref{Sharp}).

The needed dependence on $k$ appears only in the lower limit of the
integration over $q$. Therefore, we obtain
\begin{align}
k\tilde{\partial}_{k}\mathcal{L}_{1} &  =-\frac{1}{8\pi^{2}}\frac{\left(
k^{2}-d_{k}^{2}+\tilde{m}_{k}^{2}\right)  }{d^{2}}\nonumber\\
&  \times\int_{0}^{2\pi}\frac{d\phi}{2\left(  \frac{k^{2}-d_{k}^{2}+\tilde
{m}_{k}^{2}}{2kd_{k}}\right)  ^{2}+1+\cos2\phi}.
\end{align}
Once the simple integration is performed, we get
\begin{align}
& k\tilde{\partial}_{k}\mathcal{L}_{1}\nonumber\\
& =-\frac{\left(  k^{2}-d_{k}^{2}+\tilde{m}_{k}^{2}\right)  }{4\pi d_{k}^{2}%
}\frac{1}{\sqrt{\left[  2\left(  \frac{k^{2}-d_{k}^{2}+\tilde{m}_{k}^{2}%
}{2kd_{k}}\right)  ^{2}+1\right]  ^{2}-1}}.
\end{align}

Similarly, the scale derivative of $\mathcal{L}_{2}$ is given by
\begin{equation}
k \tilde{\partial}_{k} \mathcal{L}_{2} = \left[  1 + \frac{1}{2d^{2}} \left(
k^{2} -d^{2} +\tilde{m}^{2} \right)  \right]  k \tilde{\partial}_{k}
\mathcal{L}_{1} + \frac{k^{2}}{4\pi d^{2}}.
\end{equation}

To find the RG running of $\mathcal{L}_{3}$ we first note that
\begin{align}
\text{det}\left(  \frac{\slashed{\alpha}_{k}-i\tilde{m}_{k}}{\alpha_{k}%
^{2}+\tilde{m}_{k}^{2}}\right)   &  =-\frac{1}{\alpha_{k}^{2}+\tilde{m}%
_{k}^{2}},\\
\text{det}\left(  \frac{\slashed{\beta}_{k}+i\tilde{m}_{k}}{\beta_{k}%
^{2}+\tilde{m}_{k}^{2}}\right)   &  =-\frac{1}{\beta_{k}^{2}+\tilde{m}_{k}%
^{2}},
\end{align}
and
\begin{align}
&  \mathcal{X}\left(  \frac{\slashed{\alpha}_{k}-i\tilde{m}_{k}}{\alpha
_{k}^{2}+\tilde{m}_{k}^{2}},\frac{\slashed{\beta}_{k}+i\tilde{m}_{k}}%
{\beta_{k}^{2}+\tilde{m}_{k}^{2}}\right) \nonumber\\
&  =\frac{\tilde{m}_{k}^{2}-\alpha_{1k}\beta_{1k}-\alpha_{2k}\beta_{2k}%
}{\left(  \alpha_{k}^{2}+\tilde{m}_{k}^{2}\right)  \left(  \beta_{k}%
^{2}+\tilde{m}_{k}^{2}\right)  }.
\end{align}
Therefore,
\begin{align*}
&  \ -\text{det}\left(  \frac{\slashed{\alpha}_{k}-i\tilde{m}_{k}}{\alpha
_{k}^{2}+\tilde{m}_{k}^{2}}\right)  -\text{det}\left(  \frac
{\slashed{\beta}_{k}-i\tilde{m}_{k}}{\beta_{k}^{2}+\tilde{m}_{k}^{2}}\right)
\\
&  +2\mathcal{X}\left(  \frac{\slashed{\alpha}_{k}-i\tilde{m}_{k}}{\alpha
_{k}^{2}+\tilde{m}_{k}^{2}},\frac{\slashed{\beta}_{k}+i\tilde{m}_{k}}%
{\beta_{k}^{2}+\tilde{m}_{k}^{2}}\right)
\end{align*}%
\begin{equation}
=\frac{4\tilde{m}_{k}^{2}+\left(  \alpha_{1k}-\beta_{1k}\right)  ^{2}+\left(
\alpha_{2k}-\beta_{2k}\right)  ^{2}}{\left(  \alpha_{k}^{2}+\tilde{m}_{k}%
^{2}\right)  \left(  \beta_{k}^{2}+\tilde{m}_{k}^{2}\right)  }.
\end{equation}
For the sharp cutoff (\ref{Sharp}), for which $\alpha_{1k}-\beta_{1k}%
=-2id_{k}$ and $\alpha_{2k}-\beta_{2k}=0$, the last expression reads
\begin{equation}
\frac{4\left(  \tilde{m}_{k}^{2}-d_{k}^{2}\right)  }{\left(  q^{2}-d_{k}%
^{2}+\tilde{m}_{k}^{2}\right)  ^{2}+4q_{1}^{2}d_{k}^{2}}.
\end{equation}
Through the insertion of the result into Eq.(\ref{TFL3}) we obtain
\begin{equation}
\mathcal{L}_{3}=\frac{1}{\pi^{2}}\int_{k}^{\Lambda}dq\,q\int_{0}^{2\pi}%
d\phi\,\frac{\left(  \tilde{m}_{k}^{2}-d_{k}^{2}\right)  }{\left(  q^{2}%
-d_{k}^{2}+\tilde{m}_{k}^{2}\right)  ^{2}+4q^{2}d_{k}^{2}\cos^{2}\phi}.
\end{equation}

From the definition it follows that
\begin{equation}
k \tilde{\partial}_{k} \mathcal{L}_{3} = \frac{4 \left(  \tilde{m}^{2}_{k} -
d^{2}_{k} \right)  }{\left(  k^{2} - d^{2}_{k} + \tilde{m}^{2}_{k} \right)
^{2}} k \, \tilde{\partial}_{k} \mathcal{L}_{1}.
\end{equation}

\section{Canonical dimensions}

\label{AppendixC}

We discuss the dimensionality of the various quantities introduced. The
lengths, $x$, have dimensions $[x]=-1$ , i.e. the dimension of inverse
momenta. The dimension of the spinor fields, $\bar{\psi}$, $\psi$, follows
from inspection of the standard kinetic term, $\int d^{D} x \, \bar{\psi}
\partial_{x} \psi$. Since this is a contribution to the action it must be
dimensionless that yields
\begin{equation}
\left[  \bar{\psi} \right]  = \left[  \psi\right]  = \frac12 \left(  D-1
\right)  .
\end{equation}
For the same reason $[m]=1$, $[d]=1$ and $[\lambda]=2-D$.

\section{The electrostatic model}

\label{AppendixD}

To introduce the duality mapping we replace \cite{Nagaosa} in the partition
function (\ref{PHIMOD})
\begin{equation}
e^{\beta\tilde{J}\cos\Phi_{ij}}\rightarrow e^{\beta\tilde{J}}\sum_{m=-\infty
}^{m=+\infty}\exp\left[  -\frac{\beta\tilde{J}}{2}\left(  \Phi_{ij}-2\pi
m\right)  ^{2}\right]  ,
\end{equation}
where $\Phi_{ij}=\varphi_{i}-\varphi_{j}-\alpha_{ij}$, and use the Poisson sum
formula which states
\begin{equation}
\sum_{m=-\infty}^{+\infty}f(m)=\sum_{l=-\infty}^{+\infty}\int_{-\infty
}^{+\infty}d\phi\,f(\phi)e^{2\pi il\phi}. \label{Poisson_Sum}%
\end{equation}
This yields
\begin{align}
&  e^{\beta\tilde{J}\cos\Phi_{ij}}\nonumber\\
&  \rightarrow\frac{1}{\sqrt{2\pi\beta\tilde{J}}}\sum_{l_{ij}=-\infty
}^{+\infty}e^{\beta\tilde{J}}\exp\left[  -\frac{l_{ij}^{2}}{2\beta\tilde{J}%
}+il_{ij}\Phi_{ij}\right]  .
\end{align}

After substitution the result into (\ref{PHIMOD}) and omitting nonessential
factors we get
\begin{align}
\mathcal{Z}  &  =\int\mathcal{D}\varphi\,\sum_{\left\{  l_{ij}\right\}  }%
\exp\biggl(-\sum_{\langle ij\rangle}\left[  \frac{l_{ij}^{2}}{2\beta\tilde{J}%
}-il_{ij}\Phi_{ij}\right] \nonumber\\
&  \;\;\;\;+\beta h\sum_{i}\cos\varphi_{i}\biggr).
\end{align}

We now define a vector field $l_{\mu}(\boldsymbol{r})$ ($\mu=1,2$) that is
directed from the starting point $\boldsymbol{r}$, which is the left-hand side
or lower side of the link between the sites $i$ and $j$, to the other side of
the link. The vector field takes the value $l_{ij}$ on the link. The partition
function is then just the sum over all possible values $l_{\mu}(\boldsymbol{r}%
)$ of the form
\begin{align}
\mathcal{Z}  &  =\sum_{\left\{  l_{\mu}(\boldsymbol{r})\right\}  }%
\int\mathcal{D}\varphi\,\exp\biggl(-\sum_{\boldsymbol{r},\mu}\biggl[\frac
{l_{\mu}^{2}(\boldsymbol{r})}{2\beta\tilde{J}}\nonumber\\
&  -il_{\mu}(\boldsymbol{r})\left\{  \varphi(\boldsymbol{r})-\varphi
(\boldsymbol{r}+\boldsymbol{a}_{\mu})\right\}  +il_{\mu}(\boldsymbol{r}%
)\alpha_{\mu}\biggr]\nonumber\\
&  +\beta h\sum_{\boldsymbol{r}}\cos\varphi(\boldsymbol{r})\biggr),
\end{align}
where $\alpha_{\mu}$ coincides with $\alpha_{ij}$ on the $ij$-link, and
$\boldsymbol{a}_{\mu}$ is the lattice unit.

To evaluate the sum we shall make use
\begin{align}
&  \sum_{\boldsymbol{r},\mu}l_{\mu}(\boldsymbol{r})\left\{  \varphi
(\boldsymbol{r})-\varphi(\boldsymbol{r}+\boldsymbol{a}_{\mu})\right\}
\nonumber\\
&  =\sum_{\boldsymbol{r},\mu}\left\{  l_{\mu}(\boldsymbol{r})-l_{\mu
}(\boldsymbol{r}-\boldsymbol{a}_{\mu})\right\}  \varphi(\boldsymbol{r})
\end{align}
to transform the partition function into
\begin{align}
\mathcal{Z}  &  =\sum_{\left\{  l_{\mu}(\boldsymbol{r})\right\}  }%
\int\mathcal{D}\varphi\,\exp\biggl(-\sum_{\boldsymbol{r},\mu}\biggl[\frac
{l_{\mu}^{2}(\boldsymbol{r})}{2\beta\tilde{J}}\nonumber\\
&  -i\left\{  l_{\mu}(\boldsymbol{r})-l_{\mu}(\boldsymbol{r}-\boldsymbol{a}%
_{\mu})\right\}  \varphi(\boldsymbol{r})+il_{\mu}(\boldsymbol{r})\alpha_{\mu
}\biggr]\nonumber\\
&  +\beta h\sum_{\boldsymbol{r}}\cos\varphi(\boldsymbol{r})\biggr).
\end{align}

We wish to perform integration over $\varphi(\boldsymbol{r})$ from $0$ to
$2\pi$. The goal is easily accomplished with the aid of the Jacobi-Anger
expansion
\begin{equation}
e^{z \cos\varphi} = I_{0}(z) + 2 \sum_{k=1}^{\infty} I_{k}(z) \cos(k\varphi),
\end{equation}
where $I_{k}(z)$ is the modified Bessel function of the first kind.

The $\varphi$ integrals can be then done immediately that reduces the
partition function to a sum over the bond variables $l_{\mu}(\boldsymbol{r})$
with a set of $\delta$-functions restricting these variables at every site
\begin{align}
\mathcal{Z}  &  \propto\sum_{\left\{  l_{\mu}(\boldsymbol{r})\right\}  }%
\exp\left[  -\sum_{\boldsymbol{r}\mu}\left(  \frac{l_{\mu}^{2}(\boldsymbol{r}%
)}{2\beta\tilde{J}}+il_{\mu}(\boldsymbol{r})\alpha_{\mu}\right)  \right]
\nonumber\\
&  \times\prod_{\boldsymbol{r}}\biggl[\sum_{\kappa(\boldsymbol{r})=-\infty
}^{\infty}I_{\kappa(\boldsymbol{r})}(\beta h)\nonumber\\
&  \times\delta\left[  \sum_{\mu}\left(  l_{\mu}(\boldsymbol{r})-l_{\mu
}(\boldsymbol{r}-\boldsymbol{a}_{\mu})\right)  -\kappa(\boldsymbol{r})\right]
\biggr]. \label{Zvial}%
\end{align}
A presence of the magnetic field violates the "zero divergence" condition,
\begin{equation}
\sum_{\mu}\left(  l_{\mu}(\boldsymbol{r})-l_{\mu}(\boldsymbol{r}%
-\boldsymbol{a}_{\mu})\right)  =0, \label{ZeroCond}%
\end{equation}
giving the effective integer-valued charges $\kappa(\boldsymbol{r})$ confined
to lattice sites.

To gain an insight into the nature of the constraints imposed by the $\delta
$-functions it is worthy of note that $l_{\mu}(\boldsymbol{r})$ can be split
into the longitudinal and transverse parts \cite{Polyakov}
\begin{align}
l_{\mu}(\boldsymbol{r})  &  =m(\boldsymbol{r})-m(\boldsymbol{r}+\boldsymbol{a}%
_{\mu})+\sigma(\boldsymbol{r})-\sigma(\boldsymbol{r}+\boldsymbol{a}_{\mu
})\nonumber\\
&  +\varepsilon_{\mu\nu}\left[  n(\boldsymbol{r})-n(\boldsymbol{r}%
-\boldsymbol{a}_{\nu})\right]  ,
\end{align}
where $\varepsilon_{\mu\nu}$ is the standard antisymmetric tensor, the
$m(\boldsymbol{r})$ and $n(\boldsymbol{r})$ are integers, $|\sigma
(\boldsymbol{r})|<1$.

The transversal vector field $\boldsymbol{n}(\boldsymbol{r})=n(\boldsymbol{r}%
)\boldsymbol{e}_{3}$, where $\boldsymbol{e}_{3}$ is perpendicular to the plane
of the system, realizes the discrete version of the equation $\boldsymbol{l}%
(\boldsymbol{r})=\text{rot}\,\boldsymbol{n}(\boldsymbol{r})$,
\begin{equation}%
\begin{array}
[c]{c}%
l_{1}(\boldsymbol{r})=n(\boldsymbol{r})-n(\boldsymbol{r}-\boldsymbol{a}%
_{2}),\\
l_{2}(\boldsymbol{r})=-n(\boldsymbol{r})+n(\boldsymbol{r}-\boldsymbol{a}_{1}),
\end{array}
\label{Rotor}%
\end{equation}
that ensure that the "zero divergence" condition (\ref{ZeroCond}) is properly satisfied.

The $\delta$-function condition in Eq.(\ref{Zvial}) can, in turn, be satisfied
if the longitudinal part obeys the discrete Poisson equation
\begin{gather}
-\sum_{\mu}[m(\boldsymbol{r}+\boldsymbol{a}_{\mu})+m(\boldsymbol{r}%
-\boldsymbol{a}_{\mu})-2m(\boldsymbol{r})\nonumber\\
+\sigma(\boldsymbol{r}+\boldsymbol{a}_{\mu})+\sigma(\boldsymbol{r}%
-\boldsymbol{a}_{\mu})-2\sigma(\boldsymbol{r})]=\kappa(\boldsymbol{r}).
\label{Poisson}%
\end{gather}
Here, the $m(\boldsymbol{r})$ are required to be integer-valued and
$\sigma(\boldsymbol{r})$ are adjusted to keep Eq. (\ref{Poisson}).

Given that we are primary focusing on a role of the DM interaction we restrict
ourselves to the case of vanishing magnetic fields, $\beta h \to0$, when
$I_{\kappa(\boldsymbol{r})} (\beta h)$ can be replaced by the delta symbol
$\delta_{\kappa,0}$ and the trivial solution $m(\boldsymbol{r})+\sigma
(\boldsymbol{r})=0$ can be taken for Eq.(\ref{Poisson}).

By substituting (\ref{Rotor}) in Eq.(\ref{Zvial}) and taking account of
$\alpha_{\mu}=\alpha\delta_{\mu,1}$ we find
\begin{align}
\mathcal{Z} &  =\sum_{\left\{  n(\boldsymbol{r})\right\}  }\exp\biggl[-\sum
_{\boldsymbol{r}\mu}\frac{1}{2\beta\tilde{J}}\left\{  n(\boldsymbol{r}%
)-n(\boldsymbol{r}-\boldsymbol{a}_{\mu})\right\}  ^{2}\nonumber\\
&  -i\alpha\sum_{\boldsymbol{r}}\left\{  n(\boldsymbol{r})-n(\boldsymbol{r}%
-\boldsymbol{a}_{2})\right\}  \biggr].
\end{align}

Rewriting the sum running over integers $n(\boldsymbol{r})$ through the
Poisson formula (\ref{Poisson_Sum}) one obtain
\begin{align}
\mathcal{Z}  &  =\int\mathcal{D}\phi\,\sum_{\left\{  q(\boldsymbol{r}%
)\right\}  }\,\exp\biggl[-\frac{1}{2\beta\tilde{J}}\sum_{\boldsymbol{r},\mu
}\left(  \hat{\Delta}_{\mu}\phi\right)  ^{2}\nonumber\\
&  -i\alpha\sum_{\boldsymbol{r}}\hat{\Delta}_{2}\phi+2\pi i\sum
_{\boldsymbol{r}}q(\boldsymbol{r})\phi(\boldsymbol{r})\biggr],
\end{align}
where the lattice difference is defined as $\hat{\Delta}_{\mu}\phi
(\boldsymbol{r})=\phi(\boldsymbol{r})-\phi(\boldsymbol{r}-\boldsymbol{a}_{\mu
})$.

Making use of the parallel translation in the functional space
\cite{Yaglom1960}, $\phi(\boldsymbol{r})\rightarrow\phi(\boldsymbol{r}%
)-i\alpha\beta\tilde{J}x_{2}$, and carrying out Gaussian integration over
$\phi(\boldsymbol{r})$ we are led to
\begin{align}
\mathcal{Z}_{\text{eff}}  &  \propto\sum_{\left\{  q(\boldsymbol{r})\right\}
}\exp\biggl[-2\pi^{2}\beta\tilde{J}\sum_{\boldsymbol{r},\boldsymbol{r}%
^{\prime}}q(\boldsymbol{r})G(\boldsymbol{r}-\boldsymbol{r}^{\prime
})q(\boldsymbol{r}^{\prime})\nonumber\\
&  +2\pi\alpha\beta\tilde{J}\sum_{\boldsymbol{r}}q(\boldsymbol{r}%
)\,x_{2}\biggr).
\end{align}
The lattice Green function takes the form \cite{Jose1977}
\begin{align}
G\left(  \boldsymbol{r}-\boldsymbol{r}^{\prime}\right)   &  =\int\frac{d^{2}%
k}{(2\pi)^{2}}\,\frac{e^{i\boldsymbol{k}\cdot\left(  \boldsymbol{r}%
-\boldsymbol{r}^{\prime}\right)  }}{4-2\cos k_{x}-2\cos k_{y}}\nonumber\\
&  \approx-\frac{1}{2\pi}\ln\left(  \frac{\left\vert \boldsymbol{r}%
-\boldsymbol{r}^{\prime}\right\vert }{a}\right)  -\frac{1}{4}+G(0),
\end{align}
where the last term does not contain divergent terms. Interpreting
$q(\boldsymbol{r})$ as an electric charge at the position $\boldsymbol{r}$ and
the logarithmic potential as the Coulomb potential in two dimensions, the term
with $G(0)$ disappears if the charge neutrality condition, $\sum
_{\boldsymbol{r}}q(\boldsymbol{r})=0$, is imposed. The remaining part of
$G\left(  \boldsymbol{r}-\boldsymbol{r}^{\prime}\right)  $ leads to the result
(\ref{QGPF}).

\end{document}